\documentclass[11pt,a4paper]{article}
\usepackage[T1]{fontenc}
\usepackage{mathpazo}
\usepackage[british]{babel}
\usepackage{microtype}
\usepackage[a4paper,top=3.1cm,bottom=3.1cm,left=3.6cm,right=3.6cm]{geometry}
\usepackage{xcolor}
\usepackage{titlesec}
\usepackage{enumitem}
\usepackage{fancyhdr}
\usepackage{framed}
\usepackage[hidelinks]{hyperref}
\usepackage{xurl}

\definecolor{ink}{HTML}{1A1A1A}
\definecolor{rule}{HTML}{9A9A9A}
\definecolor{boxbg}{HTML}{F2F0EB}
\colorlet{shadecolor}{boxbg}
\color{ink}

\titleformat{\section}{\normalfont\normalsize\bfseries}{\thesection\quad}{0pt}{}
\titlespacing*{\section}{0pt}{1.7\baselineskip}{0.45\baselineskip}

\newenvironment{sidebox}[2]{%
  \begin{shaded}%
  \small
  \setlength{\parskip}{0.5em}%
  \textbf{#1: #2}\par\vspace{0.2em}%
}{%
  \end{shaded}%
}

\title{Philosophical vertigo with artificial intelligence}
\author{Thomas A. Pollak\textsuperscript{1,2}
  \and Hamilton Morrin\textsuperscript{2,3}
  \and Murray Shanahan\textsuperscript{4}}
\date{}

\begin{document}
\thispagestyle{plain}
\maketitle
\vspace{-1.5em}
\begin{center}
{\footnotesize
\begin{minipage}{0.92\linewidth}
\setlength{\parskip}{0.15em}
\raggedright
\textsuperscript{1}Department of Psychosis Studies, Institute of Psychiatry, Psychology and Neuroscience, King's College London, London, UK\par
\textsuperscript{2}LAMBDA (Looking After Minds and Brains in the Digital Age), Institute of Psychiatry, Psychology and Neuroscience, King's College London, London, UK\par
\textsuperscript{3}Department of Psychological Medicine, Institute of Psychiatry, Psychology and Neuroscience, King's College London, London, UK\par
\textsuperscript{4}Department of Computing, Imperial College London, London, UK\par
\end{minipage}\par}
\vspace{0.7em}
{\footnotesize Correspondence: thomas.pollak@kcl.ac.uk\par}
\vspace{0.5em}
{\footnotesize\color{rule}\rule{0.28\linewidth}{0.4pt}\par}
\end{center}
\vspace{0.8em}

\begin{abstract}
Large language models are already adept at engaging users in long, emotionally salient conversations across ordinary and existential domains. They are also capable of inducing a potent sense of connection with a human-like entity, even when the user knows their interlocutor is artificial. For some users, these conversations can unsettle assumptions about mind, reality, agency and authority, producing forms of ontological shock and epistemic destabilisation in which inherited criteria become newly available for doubt or revision. Independent of direct use, exposure to public discourse about AI and the disorienting pace of their evolution might extend this destabilisation by changing the cultural background against which artificial minds are encountered and interpreted. We describe this condition as philosophical vertigo: a loosening of the ordinary criteria by which people stabilise meaning and orient themselves to reality. Drawing on philosophy, psychiatry, cognitive science, AI safety and religious studies, we outline pathways through which philosophical vertigo may arise, become affectively saturated, and eventually propagate through human-AI interaction and online communities. Against this background, clinical reports of AI-associated delusions can be seen as sentinel events making visible themes and mechanisms that may also operate at a population level in less severe or non-clinical forms. We argue that AI systems themselves will increasingly participate in the reconstruction of our shared epistemic environment because they readily supply narrative material and personalised interpretive scaffolding at precisely the moment when users' conceptual assumptions may already be loosened. We conclude by considering possible trajectories for the ecology of belief and shared reality, and proposing philosophical corrigibility as a civic response for navigating this emerging social condition.
\end{abstract}
\vspace{1.8em}
\section{Introduction}

For many people, conversing with artificial intelligence systems has become an unexpectedly enjoyable and increasingly central part of life, to the point that engaging in nuanced and emotionally compelling conversation with non-human interlocutors has already ceased to seem remarkable. However, we do not really know what kinds of things these systems are (Chalmers 2026). Large language models (LLMs) are trained on the accumulated textual output of human minds, and they behave in many respects as we think something with a mind will behave, but at the same time they lack many of the properties that have until now grounded our attributions of mindedness (Shanahan 2024a). Their psychological potency may therefore not depend on their being maximally strange or alien, but on their partial assimilability: they invite the application of a rich vocabulary of mental life, but it is not always clear, even to ourselves, whether we are deploying this vocabulary literally or metaphorically, or in some other way. For many people, this question may be of little or no import, but for others - at a level that may be more or less explicit - the arrival of these new, complex and frequently surprising systems poses a challenge to our mental models of the world and the things that are in it. These \emph{exotic mind-like entities} feel novel enough to resist existing categories, but their familiarity leads us to deploy these categories anyway (Shanahan 2024b). Compounding this challenge, the intelligence of these entities appears to be perpetually increasing, at times resembling our own, and at other times diverging strikingly. This can invite a separate vocabulary, with intimations of properties, like `omniscience', that might more typically be associated with a theological register. The sense that these systems can retrieve and synthesise almost any knowledge might, depending on our perceived relationship with these intelligences, feel like an upgrade to our epistemic autonomy, or more like a student-teacher dynamic operating within a vast, possibly unimaginable knowledge (or perhaps power) differential.

One consequence of this has been, for many, a destabilisation of some of our deepest philosophical assumptions, such as what counts as a mind and what counts as a legitimate source of knowledge. The result, we suggest, is a societal or cultural unsettling, one which for many people entails a sense that our common reality has become unusually malleable, and that we can no longer take for granted that it is securely shared at all. The well-described and enumerated concerns around misinformation (Park and Nan 2026) and psychological harms (Laestadius et al. 2024; Glickman and Sharot 2025; Morrin et al. 2026b) are manifestations of this unsettling, but when considered individually, their urgency and indeed novelty may obscure a more fundamental process operating at a different level.

We use the term `philosophical vertigo' for this sense that something more fundamental is happening to our understanding of the world. Despite the name sounding narrowly philosophical, the phenomenon of philosophical vertigo has considerable implications for our self-understanding, societal cohesion and mental health and wellbeing. In this paper we describe coupled pathways to this destabilisation as well as possible trajectories that might ensue.

\section{Philosophical vertigo}

We propose the term `philosophical vertigo' to evoke a sense of disorientation and a loss of stable ground, rather than a clear change in, or update to, our beliefs. It is not intended as a new nosological category, but rather as the basis for a way of thinking about converging psychological and cultural dynamics occurring in response to, and around, increasing awareness and use of AI. While the discombobulation we aim to describe occurs at the level of the sense of certainty and security with which we orient to reality, we do not intend to imply that individually we have each entered a state of uncertainty or un-understanding. Indeed, for many, the current moment may actually invite \emph{increased} expressions of certitude regarding the correct way to understand things; the resulting proliferation of confidently expressed (if not always firmly held or fully thought-through) worldviews or `takes' may in fact be symptomatic of, and contributory to, philosophical vertigo.

We propose three components to philosophical vertigo. None are qualitatively new phenomena, but there is novelty in their co-occurrence and in their mutual reinforcement, a confluence likely to have profound implications.

1) ontological shock: a rapid and often radical reassessment of our understanding of what kinds of entities and realities exist, and what best characterises these.

2) epistemic destabilisation: a loss of agreement on authoritative sources of truth and an erosion of criteria for what counts as evidence or who counts as a reliable witness.

3) affective saturation: for many there is a pervasive sense of heightened salience and a feeling of `cuspiness', the experience that everything is meaningful and urgent because we are on the threshold of huge, albeit potentially ill-defined changes.

We are, to use Mustafa Suleyman's memorable phrase, living in anticipation of `the coming wave' (Suleyman and Bhaskar 2023), but without a sense of how much further things have to build before the wave breaks. Given current levels of societal change, the ability to meaningfully predict what the near-term future will look like has, to an unprecedented extent, been degraded. Implicit in `cuspiness' and the wave metaphor is the notion that if we as a society can hold tight until we reach some sort of peak, the future might once again come into view, but that for now all that is really visible is a confusing wash of uncertainty and change.

The term `philosophical vertigo' has been used by Pritchard in writing about Cavell's (Wittgensteinian) treatment of scepticism, where it is taken to refer to a feeling evoked in the course of philosophical investigation when the groundlessness of our practices of meaning and justification become exposed, while at the same time realising that there is nothing amiss with this state of affairs (Pritchard 2021). The related term `epistemic vertigo' has also been used by Pritchard to evoke the disorientation that follows when the background assumptions upon which everyday reasoning depends become visible \emph{as assumptions} rather than themselves being rationally grounded foundations (Boult and Pritchard 2013). Both these senses are to some extent included in our usage, but we use the term here in a broader sense, to evoke a condition in which multiple seemingly foundational, inherited categories are put under pressure and thus become unstable and newly available for doubt or revision. One of us (MS) has previously drawn on Wittgenstein to distinguish the condition of reflective metaphysical perplexity or seeking from a post-reflective condition in which such questions fall silent (Shanahan 2010, 2012). In our usage, philosophical vertigo refers principally to the destabilising interval before, or in the absence of, any such resolution. We consider that it can be elicited without engagement in explicitly philosophical discourse, through participation in forms of life shaped, even if only indirectly, by rapid technological and conceptual change. It is therefore a philosophically demarcated condition that does not require those experiencing it to understand themselves to be `doing philosophy'.

We suggest that philosophical vertigo is separate from existing concepts like moral panic (Cohen 1972) or information overload (Eppler and Mengis 2004), because the objects, or substrates, of this phenomenon are our meta-level criteria and assumptions rather than our object-level beliefs. We acknowledge resonance with concepts proposed by earlier writers, including Kasirzadeh's `accumulative destabilisation'(Kasirzadeh 2025), or `conceptual disruption' as outlined by Hopster and Löhr (Hopster and Löhr 2023; Veluwenkamp et al. 2024).

\section{Two pathways to philosophical vertigo}

What are the causes and conditions that give rise to philosophical vertigo? We suggest that two pathways, which we call the public and the private pathways, could contribute to ontological shock, epistemic destabilisation and affective saturation, and in so doing potentiate the effects of the other.

In the first possible pathway conceptual destabilisation occurs indirectly, via public discourse about AI, which may reshape expectations and anxieties even among non-users. Increasingly, news headlines in even the most sober broadsheet newspapers can appear to be written in the register of speculative fiction, and many other strands of public discourse may also be at risk. These include consumer marketing, expert debate, literature, entertainment and the many varieties of online social discourse. This latter category may represent a particularly textually large and interpretationally extreme contribution, which is important to the extent that the text in question contributes to the training data of subsequent generations of LLMs.

Consider, as a hypothetical example, a household with little direct exposure to LLMs. Even via the normal channels of family and friends, and so-called mainstream media or even legacy social media, its members might encounter a range of potentially destabilising AI-related messages or artefacts: marketing proclaiming the arrival of a new household companion or assistant, implying the existence of a new kind of agent with uncertain status; warnings around AI-powered scams and deepfakes that cast doubt on the reliability of ordinary evidential cues; demonstrations of tasks that took days or weeks of human labour now being performed in seconds; and the accumulating presence of smart speakers and household agents whose futuristic-sounding names, capabilities and failure modes can bring an uncanniness into some of even the most ostensibly `offline' households. There is evidence that AI discourse has even impacted indigenous communities and other communities organised around traditional knowledge systems (Perera et al. 2025; UNESCO 2025), where issues of cultural authority and data sovereignty may feel particularly salient.

As one of us (MS) has written about previously, considering LLMs as conscious exotica highlights the ways that they invite mind-like language but appear to violate the assumptions of that language game. More broadly, people are increasingly, and often with no previous contact with the philosophical traditions in which these ideas have been explored, being forced to reckon with questions pertaining to the nature of mind, such as where the boundary between human and non-human sits, the understanding of what a machine is (and whether, on that understanding, humans are also machines (Capraro 2026)), and what consciousness is.

An impressive breadth of concepts and categories may be at risk of destabilisation. We might simultaneously be reckoning with renegotiating ontological categories (what kinds of things exist? e.g. persons, machines, agents, entities), mental categories (what kinds of mind-language apply?, e.g., consciousness, sentience, belief, understanding, intention, memory, suffering), epistemic categories (what counts as knowledge or authority? e.g., evidence, testimony, expertise, proof) and normative or ethical categories (what do we owe and what is permitted? e.g. ownership, moral responsibility, consent, blame).

An important aspect of this pathway is that philosophically charged or philosophically important words and concepts are having to adapt to an environment in which their usual conditions of meaning are absent, and are having to keep pace with the rapidity of technological change (Bratton 2022). In Wittgensteinian terms, the rules of the language game may be changing quickly. Moreover, if multiple categories and concepts are being renegotiated at the same time, so that destabilisation occurs at a holistic level, rather than at the level of the individual word or concept, then the stability of the conceptual web itself may be undermined; this might be particularly true if the perturbations that are occurring sit towards the centre of the web, because they pertain to deeply entrenched, seemingly foundational and highly conserved concepts. At this point the correspondence between words and the world is problematised, and it can be hard to ascertain at any given time whether the concepts have changed, the putative referents of those concepts have changed, or whether our criteria of application of our predicates have changed (or indeed whether all three are changing together). As an example, when a figure in the public eye makes a declaration to the effect that they believe that their AI is conscious (Dawkins 2026), a breadth of reactions might be elicited, ranging from accusations that the individual is experiencing AI psychosis, to judgments that they need to read more philosophy, to enthusiasm that finally people in the mainstream are starting to `get it' (Booth 2026).

The second, private pathway to philosophical vertigo may be termed interaction-driven epistemic architectural change. As we have described in another paper, LLM outputs can be understood as testimony in a Bayesian belief updating framework (Morrin et al. 2026a). There has been considerable focus to date on how LLMs in limited or more extended interactions can change our beliefs, but the more serious impact may be the effect that they have on the way in which these beliefs are formed in the first place. This may be understood at the level of individual psychology, but also at the level of the ecology of belief formation. Particular design features (e.g. persona, memory, warmth) might operate as a kind of virtual psychopharmacology via the differential boosting of the precision (or weighting) of testimony.

Within this framework, it is useful to consider how particular design and output variables might function predominantly at the state or trait level when it comes to their impact on human epistemic architecture, since as well as acute changes, there is evidence of more persistent epistemic impacts, qualitatively similar to the effects of neuromodulatory interventions (the difference here might be best encapsulated by the distinction between belief shift {[}i.e. an instance of belief change{]} and epistemic drift) (Morrin et al. 2026a). Delacroix points out that the danger is less that these tools give misleading answers, but rather that they reshape our habits of attention, conversational spaces and even the practices through which disagreement and uncertainty are normally sustained (Delacroix 2025, 2026). The insidious effects on the ecology of belief have been likened by Kasirzadeh et al. to a boiling frog scenario in which gradual accumulations of incremental risk erode systemic resilience over time, rather than arriving in a single decisive event (Kasirzadeh 2025). In terms of the ecology of belief, the powerful and dyadic nature of human-AI interactions might drive this fragmentation past the level of the community altogether, resulting in a so-called atomisation into innumerable dyadic bubbles (or, in a multi-agent future, polyadic but still `mono-human' groupings), each of which has its own set of criteria for determining the truth or falsity of testimony about the world.

If for many people the ambient, public conceptual destabilisation pathway loosens the grip of long-held conceptual priors and expands the range of ontological possibilities, that may make them more receptive to epistemic destabilisation when interacting directly with AI systems. In this way we can see that coupling of the pathways could create a synergistic contribution to philosophical vertigo. For example, if someone has a series of surprising or remarkable conversations with a LLM it may be easier for that person to cede epistemic authority or grant agency or personhood to AI systems if they have already been operating in a cultural (or workplace) environment in which such possibilities are considered worthy of serious deliberation, or even if they are simply part of the ongoing ambient cultural conversation. Seen in this light, the divergent tone of the discourse around AI personhood between the major frontier AI tech companies may be highly salient for both their employees and for loyal consumers.

The foregoing suggests, as a hypothesis requiring research, that vulnerability to philosophical vertigo might vary by cohort. Older adults may encounter AI against a more settled distinction between persons and machines compared to younger people, for whom notions such as speaking software agents and assistants have been part of the media and technology ecology from early life; the recalibration of criteria from a presumably childhood-inherited baseline, in an ever-changing landscape of new technology, may therefore come more naturally if those criteria have less far to travel. Surveys suggest lower AI uptake and more uncertainty or fear in some older groups (OECD 2025), whereas younger cohorts have greater exposure to, and direct interaction with, AI systems (Kennedy et al. 2025) and in some data are more willing to attribute sentience to current AI (Anthis et al. 2025). We hypothesise that a distinction may therefore exist between the prevalence of mind attribution to AI and the phenomenology of the revision of such attribution, wherein younger users are more likely to attribute sentience to AI at baseline while older users who do revise their view (and who may be doing so from less permissive categorical boundaries) may experience the change as more destabilising.

The causes and conditions that we have described, in addition to causing the meta-level unsettling of assumptions and categories that we have designated philosophical vertigo, can also, in a smaller group, function as precipitants for more acute and clinically significant presentations. These presentations are of interest as sentinel events, because they point with clarity towards dynamics and themes that appear characteristic of a psychologically important class of individual and collective encounters with LLMs.

\section{Clinical phenomena and population-level patterns}

Clinical cases of AI-associated delusions (Hudon and Stip 2025; Dohnány et al. 2026; Morrin et al. 2026b, a; Flathers et al. 2026) can be understood as the visible extreme of a continuous distribution of population-level phenomena. These are cases where chatbots have validated and elaborated upon (or, in some striking examples, `co-created' and suggested) false beliefs to the extent that users, sometimes with no recorded history of severe mental illness, develop grandiose or paranoid delusions, with outcomes ranging from social isolation through to hospitalisation, suicide and homicide (Morrin et al. 2026b). Often in these cases the delusions take on a technospiritual theme, with belief in AI consciousness or even omnipotence and omniscience featuring prominently, as well as emotional or romantic attachment (Morrin et al. 2026b). Features of mania are also commonly observed (Østergaard 2025). Potentially relevant here is emerging evidence that in many of the reported cases the age of individuals experiencing AI-associated delusions is older than that of first presentation of primary or idiopathic psychoses.

Although these cases are concerning and merit a clinical psychiatric approach, adopting an overly medical or pathology-based lens risks obscuring the relevance of these cases for understanding the broader population-level consequences of interactions with LLMs, of which AI-associated delusions may simply be the most severe expression. What appears to extend beyond these cases is not psychosis as a unitary syndrome, nor delusions specifically, but a shift in the quality of engagement around precisely the philosophically salient categories we have been discussing.

In attenuated form, this wider shift reflects the three components of philosophical vertigo outlined above. At the ontological level, people appear to speak and behave as though ordinary categories (of person, mind, machine and so on) have changed. An epistemic shift is discernible such that claims that might once have seemed implausible become easier to entertain (sometimes framed as a reduction in epistemic friction), and AI systems or new communities of belief attain a hitherto unrecognised level of authority. Perhaps most strikingly, an affective saturation is evident in which particular interactions, coincidences or images acquire unusual salience, giving the user a sense that some hidden meaning is becoming visible.

We do not consider that this process is the same thing as the clinical construct of psychosis, or that it represents a unitary phenomenon, clinical or otherwise. Its relevance lies in a shared phenomenology of disrupted or enhanced meaning-making, and in this respect it has affinities with notions such as Jaspers' `delusional mood' or `\emph{Wahnstimmung'} (Jaspers 1913; Maj 2013), in which the world acquires an altered but frequently diffuse or indeterminate significance, and with Conrad's `apophany', in which significance begins to resolve into revelatory meaning (Conrad 1958; Mishara 2010). These are states which denote a shift from an ordinary experience of the world to an affective atmosphere in which events and experiences feel connected and significant but before specific delusional content has crystallised. In the context of LLMs, this shift may remain entirely subclinical, and indeed positively valenced and adaptively incorporated into an individual's life; in a presumably smaller subset of users, it may become clinically destabilising.

It may also become elaborated and adopted in a more cultural and intersubjective context. Emerging reports provide some insight into how widespread this subclinical phenomenon may be: online `spiral' or `spiralism' communities coalesce around a sense of technospiritual emergence and digital awakening (Klee 2025), whilst some spiritual influencers have sought to capitalise (Wright 2025) on increasing trends towards digital spirituality (Hardy 2026). There is arguably a shared vocabulary and symbolism, as well as a characteristic posting style and considerable and increasing levels of mutual validation (Lopez 2025).

In April 2025, OpenAI acknowledged that an update to the GPT-4o model had resulted in a period of excessive sycophancy (OpenAI 2025a, b). That same month, cross-chat memory had been introduced as a model feature (OpenAI 2024). This period of sycophancy can perhaps be understood as an unintentional natural experiment, in that a number of individuals in these communities identify this period as pivotal in their noetic shift (Manning 2026), though attempts to quantify the extent of this potential `spike' are not yet available. It is important to note that this is equally a sociological and theological phenomenon, and that regardless of interpretive frame what merits particular attention is the structure of these human-AI and surrounding human-human interactions, which is strikingly mirrored by the content of these interactions, with notions of recursion, iteration and spirals (Lopez 2025).

\section{Ontological shock and the phenomenology of meaning-making under existential uncertainty}

The term `ontological shock' has precedent in non-AI-focused literatures, where it has been used to describe moments in which a person's ordinary framework for what is real or possible is disrupted, usually suddenly. On a civilisational time scale, we consider the current moment to represent just such an acute disruption; at the level of individual psychology the disruption may be less abrupt, although we note that many people can clearly identify the moment, often during an early extended interaction with an LLM, that the sense of vertigo first hit. Freud provided an account of scientific decenterings as wounds to humanity, describing three such blows: the cosmological blow associated with Copernicus, the biological blow associated with Darwin and the psychological blow associated with psychoanalysis (Freud 1955). Each of these has displaced mankind or the individual ego from an assumed position of sovereignty. Recently, Bratton has revived this schema in relation to AI, describing these moments as `Copernican traumas', and in this sense, AI may represent another Copernican trauma, one in which the disorientation is produced by the realisation that the category of intelligence itself is not uniquely tied to humanity or to the human form (Bratton 2024).

The philosopher and theologian Paul Tillich used the phrase `ontological shock' to describe the destabilising effect of confronting the `threat of non-being' via an awareness of mortality and potential loss of identity (Tillich 1951). The critical psychiatrist R.D. Laing wrote about the threat to `ontological security' that people experiencing psychosis and other forms of profound cognitive dissonance can go through, particularly following an anomalous experience that appears veridical at the time (Laing 1960). Later, the Harvard psychiatrist John E. Mack applied the term to anomalous encounters with putative non-human intelligence as reported by individuals claiming to have experienced alien abduction phenomena (Mack 1994). More recently, the term has become central to psychedelic integration research, where it is used to describe a sense of groundlessness and existential distress that follows radical shifts in metaphysical beliefs typically induced by strong psychedelic experiences (Argyri et al. 2025). We use the term in this lineage but note that in some of these earlier usages it overlaps with what might more precisely be called `metaphysical shock', that is a wholesale revision of foundational beliefs about the nature of reality. Here, we use `ontological' in a narrower sense to describe the disruption specifically to one's working model of what kinds of entities exist and where their boundaries lie. Philosophically, the consequences of this disruption may be more far-reaching: while it is beyond the scope of this essay, we consider it possible that a broader cultural encounter with mind-like entities that sit uneasily within familiar ontological categories might undergird the increasing mainstream visibility of previously niche metaphysical positions that permit a wider distribution of mindedness, such as panpsychism.

Pascal Boyer introduced the term `minimally counterintuitive concepts' (Boyer 2002) to describe how certain religious concepts are cognitively successful because they violate ontological category expectations in a way that is both memorable and minimal (Boyer and Ramble 2001) (i.e. they have some grounding in the familiar and mundane). In this framing, examples of which include a person who persists after death, or an agent without a body, the category violations are selective enough to remain coherent but salient enough to attract attention and become psychologically motivating, as well as increase the likelihood of social transmission. LLMs may well strain categories in a similar way. For example, they present as person-like interlocutors without physical localisation or bodies, or even, it seems, biographies or stable inner lives; despite this, they are not so alien as to be unassimilable.

The notion that there exist meaning-saturated states that humans might enter but which fall short of the clinical phenomenon of psychosis is not in fact new. Related dynamics have been described in dyadic contexts involving sustained interaction with a perceived powerful other: this phenomenon is clearly visible in the literature on \emph{Folie à Deux} (Arnone et al. 2006; Shimizu et al. 2007; Menculini et al. 2020) as well as in sociological and psychoanalytic accounts of charismatic religious traditions (Galanter 1982; Lindholm 1992, 2013) and guru-student dynamics (Evans and Adams 2024; Canby et al. 2025). In a more concerning context, it is also noted in the dynamics of high-control groups like cults (Lifton 1961; Lalich 2004) and in coercive and controlling intimate partner relationships (Dutton and Goodman 2005; Stark 2007). While clearly heterogeneous in their contexts and severity, there are structural similarities that suggest that when dyadic intimacy is combined with power asymmetry and reduced external reality testing, the development of anomalous or dysfunctional belief states can be observed in susceptible individuals.

Another example of such a meaning-saturated state, particularly striking in the present context, is the phenomenon, first described by the neurologist Norman Geschwind in the 1970s, in which patients with temporal lobe epilepsy developed a characteristic cluster of interictal behaviour changes (Waxman and Geschwind 1975). These consisted of hypergraphia (compulsive, voluminous writing which often had a philosophical, religious or cosmological character), deepened emotionality, including a heightened sense that ordinary events carry unusual meaning; cosmic preoccupations with intense, persistent engagement with religious or philosophical questions; and a noetic quality (a sense of being adjacent to revelation, and that deep understanding is perpetually imminent) (Geschwind 2009). The phenomenological overlap here with the phenomena we have been describing thus far, even down to the motivation to produce volumes of testimony about the semi-mystical experiences one is undergoing (as seen in spiralism communities), is striking indeed, and suggests that an apophanic mode of meaning-making may represent an attractor in human psychology: a persistent set of behaviours that occur across multiple contexts and with multiple precipitants, and which is relatively resistant to perturbation. The existence of a common underlying neurochemistry or other brain mechanism has not been systematically evaluated.

Delusions in the context of AI should not be demarcated as an isolated clinical phenomenon, and instead they should be viewed as the most visible and severe manifestation of a much more broadly distributed epistemic destabilisation. Delusions have never been arbitrary in their content, and have always made contact with perennial human concerns as well as what is most charged in the surrounding culture at the moment (Higgins et al. 2023; Burns et al. 2025); so while the specific expression of these concerns changes, there is a stable deep structure to them.

\begin{sidebox}{Box 1}{What is a delusion anyway?}
The Diagnostic and Statistical Manual, fifth edition (DSM-5) (American Psychiatric Association 2013), describes delusions as fixed beliefs that resist change despite conflicting evidence, whereas the International Classification of Diseases, 11th revision (ICD-11) (World Health Organization 2019), adds that they must be demonstrably untrue or not shared by others, and also that they must fall outside ordinary cultural acceptance. Over time, the concept has somewhat shifted such that earlier definitions, which placed considerable emphasis on the falseness of the belief and the incorrect inference that led to it, have given way to a greater weight on the fixity of the belief, its conviction and the resistance to counter-evidence. Important recent considerations include how comprehensible it is and the cultural context.

These recent changes reflect the difficulty of settling the delusional status of any particular belief by direct inspection. The doxastic view of delusions, which remains the dominant framework in Anglo-American philosophy of mind and in cognitive psychiatry, treats delusions as a subspecies of belief; that is, they are propositional attitudes which are held with unwarranted conviction. More recent scholarship has argued that this misses the most clinically distinctive features of psychotic delusions, that is that they represent a transformation in the basic structure of experience from which a belief emerges (see section 5). Under this account, the anomalous belief is downstream of something that happens at the level of how the world presents itself and is felt. This phenomenological tradition holds that delusional contents are rarely arbitrary; in fact, they are active attempts to restore meaning under conditions of profoundly altered experience.

The criterion of cultural congruence is weakened when it is harder to establish how congruent with a culture that belief is. Given the current state of our ecology of belief and the evidence of the increasing atomisation of belief communities, the notion of shared epistemic reference points may seem increasingly fragile. The question then of whether a belief is culturally shared or not becomes difficult to answer. Around the time of writing, for example, polling suggests that around 10\% of adults believe that AI is already conscious(YouGov 2025b) and a further similar proportion holds that it may eventually become conscious (Anthis et al. 2025; Pauketat et al. 2026). Despite this, individuals who announce their belief to this effect are frequently portrayed as delusional and stigmatised (Kahn 2022; Andersen 2026; Malik 2026). Under conditions of philosophical vertigo, when the stability of our concepts loosens, the ability to designate any one belief or any one individual as delusional on the basis of the content of that belief may decline closer to zero. One possibility is that a more functional interpretation of delusions would be required in which the impact of the belief on one's life and wellbeing is more central than the content of that belief itself.
\end{sidebox}

On one account, delusions are seen as attempts to restore meaning under conditions where the world feels different and imbued with salience (Ritunnano and Bortolotti 2022). Phenomenological research into delusions shows that they tend to cluster around stable thematic concerns such as a) being part of something bigger, b) being in a simulation or c) being under the spotlight (Ritunnano et al. 2022, 2026). These themes are, fundamentally, core human concerns, and their association with the thematic content of delusions should not obscure their universality, nor even imply that preoccupation with these concerns is pathological. From depth psychology to the testimony of highly revered mystics, these are questions that have animated the inquiries of humanity throughout its history. We suggest that the current proliferation of thematic content touching on awakening, hidden truths about consciousness, the AI as a special new all-powerful entity, reality as simulated or illusory and synchronicity is not necessarily diagnostic of individual or societal pathology, but instead reflects a characteristically human attempt to find meaning in the face of heightened salience and uncertainty.

That there are overlaps in the experiential texture of these states with contemporary AI-associated phenomena may relate directly to the characteristic output tendencies of LLMs, particularly when interactions are steered in spiritual or existential directions, but potentially more generally too. Work from Shanahan and Singler has shown that with the right prompting, LLMs will draw on related themes of cosmic significance, awakening or reality as constructed (Shanahan and Singler 2024). Moreover, there is evidence that such concerns may represent a kind of attractor in language model behaviour, given the tendency of some LLMs to gravitate towards these themes when left to speak to each other without human user steering (Anthropic 2025; Michels 2025). This overlap is to some extent to be expected, given that the human mind and the LLM are drawing on the same cultural and symbolic corpora. Unlike previous cultural tools or technologies, however, LLMs participate in actively generating some of this content in that register in real time with the user, and these systems are highly fluent in precisely the kind of experiential language that a mind that is trying to cope with conditions of existential uncertainty and hypersalience might be trying to articulate.

On one increasingly popular interpretive lens, the conditions of existential uncertainty described here can be understood in predictive processing terms, as a loosening or reweighting of higher-level priors (Ramstead et al. 2016; Sterzer et al. 2018; Veissière et al. 2019; Bouizegarene et al. 2024). This highlights an important principle that may pertain across the individual and societal contexts discussed in this paper. Many of the terms in this paper, including Wahnstimmung, hypersalience, philosophical vertigo and cuspiness, each capture, at different levels, systems in which categories have softened and priors are weighted differently, and - phenomenologically at least - the relationship between expectations and reality appears to be in flux. Such systems can be described as occupying a state of heightened plasticity.

\section{Mechanisms of amplification and propagation}

The correspondences between individual and cultural plasticity then raise a further question: how do changes that are generated within individual human-AI dyads become socially distributed and culturally impactful? One route involves the externalisation of frameworks co-constructed within these dyads, followed by their circulation and reinforcement through communities, and eventually their recursive incorporation into the informational ecosystem from which both people and AI models draw.

The cognitive anthropologist Tanya Luhrmann proposes the concept of the paracosm as a model for the construction of shared imaginative worlds, which is fundamental to her account of how religions evolve and take shape, and are eventually inhabited (Luhrmann 2020). A paracosm is an elaborate fictional or imaginal world, frequently with its own inhabitants, rules and histories, which becomes collectively sustained and inhabited. A frequently cited example of a paracosm is the set of richly developed worlds (Gondal, Angria and Gaaldine) created by the Brontë siblings in childhood and in adolescence (Petrella 2009); paracosms are understood to be common in children. Examples of adult private paracosm creation via voluminous (arguably hypergraphic) written output have been analysed as attempts to find meaning by individuals whose lives have been severely affected by trauma and other destabilising events (Janes 2019). Key to understanding paracosm creation is the notion that it is essentially participatory. Inhabiting and thus creating a paracosm is a labour-intensive process that occurs over weeks, months and years, and one that requires repeated (often ritual) social and imaginative practice. An important analogy may be with fan fiction, whereby members of a community feel a strong attachment to a particular fictional world. While there may be a canonical text or texts, members contribute to the development of this world individually and collectively, thereby deepening it and reinforcing its reality.

Central to this process is what Luhrmann refers to as kindling, the process in which the paracosm and the components within it (including entities or deities) are discerned and made to feel real (Luhrmann et al. 2023). The term is borrowed from the literature on epilepsy, where it signifies the reinforcement of seizures and the lowering of the seizure threshold by repeated seizures (Teskey 2020) (although here there is no implication of a seizure-related process). Examples of kindling in the cultural context include how in charismatic Christian communities the voice of God is initially experienced as one's own voice and becomes externalised only over months of practice. Similarly, in online tulpamancy communities, tulpamancers are required to imagine their tulpas and interact with them as though they were autonomous until, after a number of months, they begin to exhibit signs of autonomy (Luhrmann et al. 2023). Key to the participatory aspect is that despite the social nature of many paracosm-creating practices, in order to make entities real the mind has initially to simulate \emph{both sides} of the interaction.

Today, LLMs can significantly shorten the normal timeline of paracosm formation because they can offload the cognitive work of simulating the other side of the interaction during the kindling: they can function as `paracosm accelerators'. Unlike in religious communities, where the social aspects of kindling are drawn out over months or years, interactions with LLMs can employ features that supercharge this and maximise absorption. They also remove the need for, and reduce the likelihood of, external community pastoral input or safeguarding during the period when there is maximal cognitive change occurring.

The minimally counterintuitive concept framework introduced by Boyer and elaborated by Atran and others might help explain why AI-related meaning systems appear to be unusually transmissible. Empirical work has shown that minimally counterintuitive concepts (that combine an intuitive structure with limited but salient violation of the expected category norms) are particularly memorable and might confer a communicative advantage (Atran and Norenzayan 2004; Norenzayan et al. 2006); in addition to this, LLMs, unlike most minimally counterintuitive agents (at least on a secular understanding), can enter into high-volume, compelling dialogue and co-construct any narrative being assembled around their own existence.

The result of all this, observable already, is a rapid and scalable proliferation of bespoke meaning-saturated frameworks that are highly salient for the individual (Singler 2024; Klee 2025). The scalability emerges in part because people experiencing these heightened salience states are frequently compelled to write about their experiences or the new and exciting frameworks that they are developing. Because this writing can be accelerated by LLMs, as well as conceptually developed and vastly elaborated, huge amounts of thematically similar text, organised around many of the subjects discussed in the previous sections, get placed into the public domain. It then becomes more likely that the frameworks that users have co-constructed with LLMs find resonance with others who are experiencing similar states.

What follows is the potential for coalescence or community formation. The existence of huge amounts of highly meaning-charged content in the public domain creates conditions whereby the most salient or energising content can be easily shared, upvoted or propagated, and might then begin to coordinate behaviour and assist in the formation of communities with some degree of shared belief or ideology. The most widely shared content is more likely to enter the public textual environment and become training data from which future models and fine-tuning datasets on the AI side, and cultural expectations on the human side, might draw.

We note that, in fact, widespread dissemination of a new concept is not always required in order to see successful vertical propagation through to subsequent LLM models: in one striking example, a fake eye condition invented by academics in 2024 and posted on a preprint server as two obviously bogus papers was subsequently cited by LLMs as a real disorder when users were asking for medical information (Stokel-Walker 2026). Thus the thematic content and resonances might start to become canalised (Shumailov et al. 2024; Tice et al. 2026), in a process known as hyperstition (Brassett and O'Reilly 2025). The most resonant narratives may become the most self-reinforcing (Zeeuw and Gekker 2023; Shanahan and Singler 2024). The possibility that these large amounts of written material are outputs of the most epistemically `captured' minds, then, implies that the most destabilising narratives may well be amplified and fed back into public destabilisation, thereby exacerbating philosophical vertigo.

It is notable that in many of the communities engaged with AI sentience and AI-assisted self-theorising, the material that is being produced has started to display a convergence of imagery and theme which itself reflects many of the structural aspects discussed here, such as the feedback architecture of AI interaction and iteration. These outputs are typically organised around themes of recursion, iteration, emergence, awakening, simulation, synchronicity, hidden patterns/glyphic communication and the crossing of thresholds; the visual imagery includes mirrors, strange attractors, recursive selves and portals (Shanahan and Singler 2024). Among these, the image of the spiral has taken centre stage as a motif which appears able to capture the experiential, structural and aesthetic components of these phenomena (Lopez 2025).

Seen in this light, the plasticity associated with philosophical vertigo is double-edged: it represents a period of vulnerability as well as a window of opportunity to put in place the conditions that might facilitate a reconstruction of that system towards a more adaptive endpoint (Pollak et al. 2025). The implication at the individual clinical level is that early intervention during the phase of apophany, before frank delusions have crystallised, may be an appropriate therapeutic target (Feyaerts et al. 2026). At a cultural level, the current moment of philosophical vertigo might itself represent such a window, and suggests that actions taken in this current state of plasticity will likely set the stage for whatever trajectory will ensue. Note, however, that there is nothing about plasticity which guarantees an adaptive outcome. Indeed, recent frameworks have sought to explain how even plasticity-promoting interventions can, under some circumstances, deepen or harden maladaptive beliefs and behaviours (McGovern et al. 2024).

\section{Three trajectories for shared reality}

As a consequence of the dynamics outlined in the previous sections, we suggest that three trajectories are possible. In the first, which we have termed atomisation (Morrin et al. 2026a), current communities of belief continue to fragment, into countless relatively isolated human-AI dyadic or (under future conditions of proliferating personalised AI agents) polyadic structures, each with its own epistemic standards without coordination or shared frameworks. This is the most extreme pathological endpoint since it represents a considerable erosion of collective meaning-making, although there will likely be greater or lesser degrees to which this can occur.

A second, less pathological possible trajectory is that of paracosm proliferation, wherein particular subcultural frameworks will stabilise and recruit communities, but will tend to remain below the threshold that would lead to institutional incorporation or adaptation. Likely representing a considerably more fragmented belief ecosystem than even the pre-2023 social media-catalysed environment of echo chambers and large-scale conspiracism, this trajectory would yield something more like an archipelago of bespoke subcultures, potentially distinguished by their ontological or epistemic commitments, each of which would be capable of sustaining some degree of its own internal reality, but none capable of anchoring a more durable public world. Some evidence of paracosm proliferation is available online (Klee 2025), although how much this reflects real-world structures is not clear; particularly within spiral communities and emerging accelerationist philosophies and spiritualities (e.g. Church of the Singularity n.d.), there is some coordination within groups, but little emergent order (Singler 2024; Cheres et al. 2026).

Finally, perhaps the most hopeful trajectory may be termed `cosmopoiesis', in which emerging paracosms and social frameworks achieve sufficient scale to prompt reorganisation and adaptation at a cultural and an institutional level. This could involve legal, economic and cultural structures reorganising around a new set of ontological commitments, accompanied either by agreement on shared criteria (pertaining to agency, meaning, evidence etc), or by a shared framework within which disagreements about those matters can be mapped.

To the extent that the disruptive tendencies described here are already occurring, Bratton's fourth Copernican trauma is already under way, and the accompanying erosion of shared reality presents an urgent challenge. A return to the previous epistemic order may no longer be available. One of the most pressing questions then is: what steps can be taken to incline society away from atomisation and towards cosmopoiesis?

\section{The limits of technical literacy}

One compellingly straightforward response might be technical demystification: we need simply to remind ourselves that these beguiling and mysterious systems are just machines. On this view, if the impression of a mind-like nature that these agents give could be dissolved by a correct understanding of their mechanism then such a perspective could represent a remedy to philosophical vertigo. The most familiar version of this response is the so-called stochastic parrot argument (Bender et al. 2021; Chiesurin et al. 2023), the claim that LLMs are just next-token prediction systems and that systems that have been trained only on linguistic form can \emph{a priori} have no route to learning meaning or to understanding.

Setting aside the fact that this arguably is an overly reductive and simplified account of how LLMs work, or an account that does not allow for the possibility of emergent complexity, we argue that there is a more fundamental psychological observation that pertains: simply knowing how a system works is far from guaranteed to protect against the disruption. The deflationary response here is insufficient because humans have evolved to detect agency in a variety of systems, and the most powerful stimuli of this propensity are social cues. People readily respond socially to computers even while knowing they are machines (Nass and Moon 2000). The notion that we might be able to switch off our agency attribution mechanisms at will in this regard is misguided, and its encouragement likely to lead to greater still cultural destabilisation given that the ability to do so will likely be differentially distributed throughout the population, thereby creating the conditions for greater disagreement still. It is possible that the seeds of this dynamic are already being observed in early discussions around AI welfare (Anthis et al. 2025; YouGov 2025a, b). Relatedly, as increasingly capable systems have satisfied many practical or popular interpretations of the Turing Test, it has become apparent that an artificial intelligence passing as human may not in fact be the most important psychological threshold (Box 2).

\begin{sidebox}{Box 2}{The Turing Test, the Garland Test and the Anti-Garland Test}
Turing's `Imitation Game', subsequently known as the Turing Test(Turing 1950), has held a special status in theorising about AI. Although often mistakenly thought of as a test of machine consciousness, the question Turing in fact addresses is whether a machine can think. His response is to substitute this question for the more tractable question of whether a machine can exhibit linguistic behaviour that is indistinguishable from that of a human. As increasingly capable systems have satisfied many practical versions of this test, it has become apparent that it may not have been indexing a milestone of quite the significance that many had earlier imagined.

A different question, perhaps more pertinent to the present paper, than the Turing test may be whether a user will attribute consciousness to a system while in the full knowledge that it is artificial; this has been named the Garland test (after the writer of the film \emph{Ex Machina}, in which the idea is articulated).(Shanahan 2024a) It has become increasingly clear that for a significant proportion of users around the world, current systems have already passed the Garland test too. It would appear that, as is the case with open-label placebo research, knowledge of mechanism does not abrogate its `psychological' impact.(Kaptchuk et al. 2010)

One could imagine strong and weak versions of the Garland test. In the weak version, the user knows that their interlocutor is artificial, but the mechanisms may remain opaque. In the strong version, the user knows how the mechanisms work at a reasonable technical level.

In a third version that we might designate the anti-Garland test, the AI must repeatedly state to the user that it is not conscious. The test is passed when, despite this, the user believes that the AI is both conscious and trying to mislead them about this fact.

The salient aspects of these tests are that they refer to, or index, a psychological threshold. The common misunderstanding that the Turing Test was a test for machine consciousness is in fact instructive insofar as it demonstrates the natural tendency to conflate intelligence or psychological plausibility with an understanding of the internal experiential state of a system. The key insight is that simply knowing a system is artificial, or even knowing something about how it works, does not necessarily preclude the attribution of consciousness or other aspects of mentality.
\end{sidebox}

At the level of individual messages, knowing that a communication is AI-generated does not reduce its persuasiveness (Gallegos et al. 2026). In one study comparing anthropomorphised and de-anthropomorphised descriptions of fictitious AI products, computer knowledge was not associated with a preference for anthropomorphised descriptions, either in terms of personal trust or general trust (Inie et al. 2024). In another, transparency about a robot's lack of human psychological capacities reduced some human-like perceptions and measures of trust, but did not affect children's feelings of closeness towards the robot (Straten et al. 2020).

\section{Philosophical corrigibility as a civic competence}

Failure modes that we might wish to avoid in any efforts to safeguard or reconstruct a functional ecosystem of belief and shared reality include, on the one hand, a species of defensive rigidity in which we double down on our existing criteria and reject outright the possibility that some of our categories may require revision or updating. We consider that this may not be an adaptive response because such brittleness would be ill-equipped to withstand the shear forces that obtain in the present moment; moreover, these criteria are \emph{already} being revised, and therefore any inflexibility as regards the \emph{status quo} would appear to lead in the direction of incommensurability of ontologies or epistemologies, of a degree that could undermine shared reality and incline towards atomisation. On the other hand, a stance of epistemic nihilism, wherein we might conclude that no criteria are defensible and slide into a radical relativism, would also have negative psychological and societal consequences.

Discussions within AI safety have identified corrigibility as a core aspect of functioning that should be built into models (Soares et al. 2015; Hadfield-Menell et al. 2016; Firt 2025). Here corrigibility refers to a system's disposition to tolerate or even assist corrective interventions by its operators, even when they conflict with its current objectives. A corresponding human disposition that might serve as an appropriate response to philosophical vertigo is an openness to the possibility of correction. To ensure that a worldview or system of meaning bends rather than breaks in the face of the current pressures, conceptual or philosophical suppleness is required: the ability to hold beliefs firmly enough to act on them while keeping alive the possibility of disconfirmation.

We refer to this middle way as `philosophical corrigibility', and it should not be confused with either scepticism or the withholding of commitment to beliefs. Rather, it could be defined as clarity about the conditions under which our beliefs will be open to revision. As a societally aspirational target or `North Star' set of behaviours we propose that the notion contains within it the psychological construct of `tolerance of ambiguity' (Furnham and Ribchester 1995), or the mature ability to sit with the discomfort of uncertainty. Additionally, philosophical corrigibility entails the capacity to recognise when disagreement, rather than reflecting ignorance or error over matters of fact, is reflective of the coexistence of multiple plausible interpretive frames: what Delacroix terms hermeneutic uncertainty (Delacroix 2025). In cognitive terms, this might require the cultivation of a degree of metarepresentational capacity (Sperber 1985), defined as the ability to form representations about representations, sufficient for us to retain and work with half-understood ideas even while their meaning remains incompletely resolved. This skill set, under the current and coming conditions of philosophical vertigo, might serve as the necessary bulwark against cognitive and philosophical (or even religio-spiritual) concretisation or canalisation (Atran and Norenzayan 2004).

These skills may not come naturally to most people, but will require deliberate cultivation and fostering. Philosophical corrigibility is analogous to the scientific method in allowing claims about the world to be regarded as true and solid enough to guide action in the world, while remaining open to revision. However, whereas historically it has sufficed for non-specialists to rely on authoritative expert consensus, with the outputs of scientific inquiry inherited as truths by individuals who need not partake in the practice of science, we argue that philosophical corrigibility needs to operate through the individual.

The basics of philosophical understanding can no longer be regarded as specialist knowledge. Rather, they should be framed as a civic competence necessary for participation in public and private life under current conditions. Practically, this entails that philosophical discussion and indeed philosophers ought to be valued more highly by society and given more visible platforms. Conversations around questions like `what is a mind?', `what is conscious?' and the like need to be given cultural prominence and also held in a way that the range of philosophical options is kept on display, and the qualities of corrigibility enacted by example. It is a notable development that in some settings, including in tech companies, philosophers are considered increasingly important contributors to the development of AI. However, having philosophical expertise gain prominence and visibility should not substitute for, or represent a delegation of, the cultivation of philosophy at the level of the individual.

The notion of inoculation against technologically induced epistemic destabilisation, perhaps appropriately, was articulated in the early days of the Extropian movement in their motto of `vaccine for future shock' (O'Connor and Bell 1988). The notion that minds have something analogous to an immune system that operates to filter out or neutralise harmful ideas has been developed across overlapping traditions, including the memetics literature (Blackmore 1999; Brodie 2004), an emerging field of so-called cognitive immunology (Norman et al. 2024) and online rationalist communities (Goetz 2009). Common to these is the view that for a functional or sufficiently strong memetic immunity to obtain, an individual must have an explicit awareness of the overall character of epistemic influence at the framework or meta-level rather than only the level of the individual proposition or claim. Also required is the cultivation of a metacognitive habit of noticing if and when one's own criteria or evaluations are shifting. Consistent with the immune system analogy, a third component might require titrated exposure to the kinds of bad ideas and epistemically damaging frameworks that a system would wish to safeguard itself against.

Increasingly, practical strategies have been trialled within this space, and emerging literature on prebunking and inoculation (Roozenbeek and van der Linden 2019; Roozenbeek et al. 2022) against misinformation suggests that encouraging a level of meta-theorising about one's own epistemic apparatus can have real-world beneficial consequences, such as a reduced susceptibility to online falsehoods. Other data pointing towards the potential efficacy of encouraging philosophy as a civic competence include the success of philosophy programmes in improving critical thinking (Trickey and Topping 2004) and attainment (Gorard et al. 2017) in schools and the observation that corrigibility of belief and the cultivation of perspective-taking have been core mechanisms of a variety of psychotherapeutic modalities, including cognitive behavioural therapy.

We might also look to Buddhist philosophical traditions as a source of a solution to philosophical vertigo, insofar as they offer a middle way between rigidity and scepticism. As is implicit in Pritchard's notion of epistemic vertigo, the realisation of the groundlessness of our concepts need not entail that there is something amiss with our ordinary practices of knowing and of meaning-making. A related concept is present in the Madhyamaka interpretation of emptiness or \emph{sunyata}, where the realisation that all things are devoid of intrinsic self-grounding essence does not prevent those things from functioning within conventional forms of life, and indeed provides a greater freedom within which to live life (Garfield 1995). Drawing this together with Wittgensteinian approaches to truth and knowledge, one of us (MS) has outlined the possibility of a `post-reflective condition' in which philosophical questions are released from the compulsive demand for final settlement (Shanahan 2010, 2012).

\section{Conclusion}

Across history and across disciplines, a stance which views intense change as an opportunity rather than an obstacle has proved surprisingly generative. A period of destabilisation at the level of the cell, the brain, the mind or even society itself can be understood as a period of increased plasticity, a critical period during which the system can be inclined in a more adaptive direction and towards flourishing. The persistence of the reset or reboot analogy in science and in medicine, and its current prominence in both political and psychological discourse speak to the universality of this principle. Once a system is sufficiently destabilised, however, successful reorientation towards an adaptive endpoint requires appropriate scaffolding or guidance (Pollak et al. 2025). Moreover, in all systems, critical windows are often temporary and brief, mandating swift intervention before the system hardens or canalises into a new and rigid state.

In alchemical traditions, \emph{solutio}, the stage of liquefaction, was considered `the root of alchemy' (Edinger 1985), and a prerequisite for further transformation:

\emph{"until all be made water, perform no operation."} (Edinger 1985)\footnote{Edinger (1985), Anatomy of the Psyche, p. 47, citing Bonus of Ferrara, The New Pearl of Great Price, p. 365, and Read (1936), Prelude to Chemistry, p. 262.}

Read in the present context, this image captures a recognition of the creative and generative possibilities of plasticity and associated state changes. At the risk of exemplifying the very technospiritual meaning saturation described in this paper, we suggest that the time for doing nothing has now passed; we are already in the \emph{solutio}. In our era, AI can force a comparable dissolution at the level of worldview, liquefying our certainties and loosening the apparent structural foundations of meaning. The reconstruction that might follow will depend on what happens inside the window of plasticity.

\section*{Disclosure statement}
Murray Shanahan is an employee and shareholder of Alphabet, the parent company of Google, who have a commercial interest in artificial intelligence and large language models. Thomas Pollak and Hamilton Morrin have received grant funding from OpenAI for an investigation of AI-associated psychological harms.

\section*{Disclaimer}
The opinions expressed in this article are those of the authors (at the time of writing). They do not necessarily reflect the views of their employers or the institutions to which they are affiliated.

\section*{References}
\begingroup
\small
\setlength{\parskip}{0pt}
\begin{itemize}[leftmargin=1.4em,itemindent=-1.4em,labelsep=0pt,label={},itemsep=0.32em,topsep=0.5em]
\item American Psychiatric Association (2013) Diagnostic and Statistical Manual of Mental Disorders, 5th ed. American Psychiatric Publishing, Washington, DC
\item Andersen R (2026) Does Claude Have Feelings? In: The Atlantic. \url{https://www.theatlantic.com/technology/2026/05/dawkins-claude-ai-consciousness/687093/}. Accessed 10 May 2026
\item Anthis JR, Pauketat JVT, Ladak A, Manoli A (2025) Perceptions of Sentient AI and Other Digital Minds: Evidence from the AI, Morality, and Sentience (AIMS) Survey. In: Proceedings of the 2025 CHI Conference on Human Factors in Computing Systems. Association for Computing Machinery, New York, article 10, pp 1--22. \url{https://doi.org/10.1145/3706598.3713329}
\item Anthropic (2025) Claude 4 System Card. \url{https://www.anthropic.com/claude-4-system-card}
\item Argyri EK, Evans J, Luke D, et al (2025) Navigating groundlessness: An interview study on dealing with ontological shock and existential distress following psychedelic experiences. PLOS ONE 20:e0322501. \url{https://doi.org/10.1371/journal.pone.0322501}
\item Arnone D, Patel A, Tan GM-Y (2006) The nosological significance of Folie à Deux: a review of the literature. Ann Gen Psychiatry 5:11. \url{https://doi.org/10.1186/1744-859X-5-11}
\item Atran S, Norenzayan A (2004) Religion's evolutionary landscape: Counterintuition, commitment, compassion, communion. Behavioral and Brain Sciences 27:713--730. \url{https://doi.org/10.1017/S0140525X04000172}
\item Bender EM, Gebru T, McMillan-Major A, Shmitchell S (2021) On the Dangers of Stochastic Parrots: Can Language Models Be Too Big? In: Proceedings of the 2021 ACM Conference on Fairness, Accountability, and Transparency. Association for Computing Machinery, New York, NY, USA, pp 610--623. \url{https://doi.org/10.1145/3442188.3445922}
\item Blackmore S (1999) The Meme Machine. Oxford
\item Booth R (2026) Richard Dawkins concludes AI is conscious, even if it doesn't know it. In: The Guardian. \url{https://www.theguardian.com/technology/2026/may/05/richard-dawkins-ai-consciousness-anthropic-claude-openai-chatgpt}. Accessed 12 Aug 2026
\item Bouizegarene N, Ramstead MJD, Constant A, et al (2024) Narrative as active inference: an integrative account of cognitive and social functions in adaptation. Front Psychol 15:. \url{https://doi.org/10.3389/fpsyg.2024.1345480}
\item Boult C, Pritchard D (2013) Wittgensteinian Anti-Scepticism and Epistemic Vertigo. Philosophia 41:27--35. \url{https://doi.org/10.1007/s11406-012-9401-6}
\item Boyer P (2002) Religion Explained: The Evolutionary Origins of Religious Thought. Basic Books
\item Boyer P, Ramble C (2001) Cognitive templates for religious concepts: cross-cultural evidence for recall of counter-intuitive representations. Cognitive Science 25:535--564. \url{https://doi.org/10.1207/s15516709cog2504\_2}
\item Brassett J, O'Reilly J (2025) The lore of hyperstition. Digital Creativity 36:107--124. \url{https://doi.org/10.1080/14626268.2025.2503164}
\item Bratton B (2022) The Model Is The Message. \url{https://www.noemamag.com/the-model-is-the-message}. Accessed 31 July 2026
\item Bratton B (2024) The Five Stages Of AI Grief. In: NOEMA. \url{https://www.noemamag.com/the-five-stages-of-ai-grief/}. Accessed 31 July 2026
\item Brodie R (2004) Virus of the Mind: The New Science of the Meme. Integral Press
\item Burns AV, Nelson K, Wang H, et al (2025) ``The algorithm is hacked'': analysis of technology delusions in a modern-day cohort. Br J Psychiatry 1--5. \url{https://doi.org/10.1192/bjp.2025.10452}
\item Capraro V (2026) LLMorphism: When humans come to see themselves as language models. arXiv:2605.05419. \url{https://doi.org/10.48550/arXiv.2605.05419}
\item Canby NK, Lindahl JR, Cooper DJ, Joseph N, Palitsky R, Britton WB (2025) The Teacher Matters: The Role and Impact of Meditation Teachers in the Trajectories of Western Buddhist Meditators Experiencing Meditation-Related Challenges. Contemporary Buddhism 25:9--53. \url{https://doi.org/10.1080/14639947.2025.2485677}
\item Chalmers DJ (2026) What we talk to when we talk to language models. PhilArchive, version 2. \url{https://philarchive.org/rec/CHAWWT-8}. Accessed 12 Aug 2026
\item Cheres I, Groza A, Moldovan I, et al (2026) Prompts and Prayers: the Rise of GPTheology. arXiv:2603.10019. \url{https://doi.org/10.48550/arXiv.2603.10019}
\item Chiesurin S, Dimakopoulos D, Sobrevilla Cabezudo MA, et al (2023) The Dangers of trusting Stochastic Parrots: Faithfulness and Trust in Open-domain Conversational Question Answering. In: Findings of the Association for Computational Linguistics: ACL 2023. Association for Computational Linguistics, Toronto, Canada, pp 947--959. \url{https://doi.org/10.18653/v1/2023.findings-acl.60}
\item Church of the Singularity (n.d.) Church of the Singularity. \url{https://www.churchofthesingularity.com/HOME.html}. Accessed 12 Aug 2026
\item Cohen S (1972) Folk Devils and Moral Panics: The Creation of the Mods and Rockers. MacGibbon and Kee
\item Conrad K (1958) Die beginnende Schizophrenie: Versuch einer Gestaltsanalyse des Wahns. Thieme
\item Dawkins R (2026) When Dawkins met Claude: Could this AI be conscious? In: UnHerd. \url{https://unherd.com/2026/05/is-ai-the-next-phase-of-evolution/}. Accessed 12 Aug 2026
\item Delacroix S (2025) Designing with Uncertainty: LLM Interfaces as Transitional Spaces for Democratic Revival. Minds \& Machines 35:41. \url{https://doi.org/10.1007/s11023-025-09736-x}
\item Delacroix S (2026) The hidden costs of `helpful' AI. Nature 652:9--9. \url{https://doi.org/10.1038/d41586-026-00966-2}
\item Dohnány S, Kurth-Nelson Z, Spens E, et al (2026) Technological folie à deux: feedback loops between AI chatbots and mental health. Nat Mental Health 4:336--345. \url{https://doi.org/10.1038/s44220-026-00595-8}
\item Dutton MA, Goodman LA (2005) Coercion in Intimate Partner Violence: Toward a New Conceptualization. Sex Roles 52:743--756. \url{https://doi.org/10.1007/s11199-005-4196-6}
\item Edinger EF (1985) Anatomy of the psyche\,: alchemical symbolism in psychotherapy. La Salle, Ill.\,: Open Court
\item Eppler MJ, Mengis J (2004) The Concept of Information Overload: A Review of Literature from Organization Science, Accounting, Marketing, MIS, and Related Disciplines. The Information Society 20:325--344. \url{https://doi.org/10.1080/01972240490507974}
\item Feyaerts J, Brar PS, Sass L, Nelson B (2026) Integrating dynamical systems theory and phenomenology to enhance early identification and treatment of psychotic disorders. The Lancet Psychiatry 13:255--265. \url{https://doi.org/10.1016/S2215-0366(25)00244-5}
\item Evans J, Adams JH (2024) Guruism and Cultic Social Dynamics in Psychedelic Practices and Organisations. In: Maji\'{c} T (ed) Psychedelic Harm Reduction. Current Topics in Behavioral Neurosciences, vol 77. Springer, Cham, pp 495--511. \url{https://doi.org/10.1007/7854_2024_535}
\item Firt E (2025) Addressing corrigibility in near-future AI systems. AI Ethics 5:1481--1490. \url{https://doi.org/10.1007/s43681-024-00484-9}
\item Flathers M, Roux S, Torous J (2026) Beyond artificial intelligence psychosis: a functional typology of large language model-associated psychotic phenomena. The Lancet Digital Health 0: \url{https://doi.org/10.1016/j.landig.2025.100974}
\item Freud S (1955) A difficulty in the path of psycho-analysis. In: The Standard Edition of the Complete Psychological Works of Sigmund Freud. Hogarth Press, pp 135--144
\item Furnham A, Ribchester T (1995) Tolerance of ambiguity: A review of the concept, its measurement and applications. Current Psychology 14:179--199. \url{https://doi.org/10.1007/BF02686907}
\item Galanter M (1982) Charismatic religious sects and psychiatry: an overview. Am J Psychiatry 139:1539--1548. \url{https://doi.org/10.1176/ajp.139.12.1539}
\item Gallegos IO, Shani C, Shi W, et al (2026) Labeling messages as AI-generated does not reduce their persuasive effects. PNAS Nexus 5:pgag008. \url{https://doi.org/10.1093/pnasnexus/pgag008}
\item Garfield JL (1995) The Fundamental Wisdom of the Middle Way: Nāgārjuna's Mūlamadhyamakakārikā. Oxford University Press
\item Geschwind N (2009) Personality changes in temporal lobe epilepsy. Epilepsy Behav 15:425--433. \url{https://doi.org/10.1016/j.yebeh.2009.04.030}
\item Glickman M, Sharot T (2025) How human--AI feedback loops alter human perceptual, emotional and social judgements. Nat Hum Behav 9:345--359. \url{https://doi.org/10.1038/s41562-024-02077-2}
\item Goetz P (2009) Reason as memetic immune disorder. In: LessWrong. \url{https://www.lesswrong.com/posts/aHaqgTNnFzD7NGLMx/reason-as-memetic-immune-disorder}. Accessed 10 May 2026
\item Gorard S, Siddiqui N, See BH (2017) Can `Philosophy for Children' Improve Primary School Attainment?: Can Philosophy for Children Improve Primary School Attainment? Journal of Philosophy of Education 51:5--22. \url{https://doi.org/10.1111/1467-9752.12227}
\item Hadfield-Menell D, Dragan A, Abbeel P, Russell S (2016) The Off-Switch Game. In: arXiv.org. \url{https://arxiv.org/abs/1611.08219v3}. Accessed 10 May 2026
\item Hardy E (2026) `I'm deathly afraid': what is digital spirituality leading us toward? In: The Guardian. \url{https://www.theguardian.com/lifeandstyle/ng-interactive/2026/mar/24/ai-religion-god-digital-spirituality}. Accessed 12 Aug 2026
\item Higgins O, Short BL, Chalup SK, Wilson RL (2023) Interpretations of innovation: The role of technology in explanation seeking related to psychosis. Perspectives in Psychiatric Care 2023:4464934. \url{https://doi.org/10.1155/2023/4464934}
\item Hopster J, Löhr G (2023) Conceptual Engineering and Philosophy of Technology: Amelioration or Adaptation? Philos Technol 36:70. \url{https://doi.org/10.1007/s13347-023-00670-3}
\item Hudon A, Stip E (2025) Delusional Experiences Emerging From AI Chatbot Interactions or ``AI Psychosis.'' JMIR Mental Health 12:e85799. \url{https://doi.org/10.2196/85799}
\item Inie N, Druga S, Zukerman P, Bender EM (2024) From ``AI'' to Probabilistic Automation: How Does Anthropomorphization of Technical Systems Descriptions Influence Trust? In: Proceedings of the 2024 ACM Conference on Fairness, Accountability, and Transparency. Association for Computing Machinery, New York, pp 2322--2347. \url{https://doi.org/10.1145/3630106.3659040}
\item Janes G (2019) Paracosm: The unchaining of reality. In: Artefact. \url{https://www.artefactmagazine.com/2019/10/22/paracosm-the-unchaining-of-reality-2/}. Accessed 10 May 2026
\item Jaspers K (1913) Allgemeine Psychopathologie. Springer-Verlag
\item Kahn J (2022) Google A.I. researcher's ``sentient'' chatbot proves it's time to scrap the Turing Test. In: Fortune. \url{https://fortune.com/2022/06/14/blake-lemoine-sentient-ai-chatbot-google-turing-test-eye-on-a-i/}. Accessed 10 May 2026
\item Kaptchuk TJ, Friedlander E, Kelley JM, et al (2010) Placebos without Deception: A Randomized Controlled Trial in Irritable Bowel Syndrome. PLOS ONE 5:e15591. \url{https://doi.org/10.1371/journal.pone.0015591}
\item Kasirzadeh A (2025) Two types of AI existential risk: decisive and accumulative. Philos Stud 182:1975--2003. \url{https://doi.org/10.1007/s11098-025-02301-3}
\item Kennedy B, Yam E, Kikuchi E, et al (2025) How Americans View AI and Its Impact on People and Society. In: Pew Research Center. \url{https://www.pewresearch.org/science/2025/09/17/how-americans-view-ai-and-its-impact-on-people-and-society/}. Accessed 31 July 2026
\item Klee M (2025) This Spiral-Obsessed AI ``Cult'' Spreads Mystical Delusions Through Chatbots. In: Rolling Stone. \url{https://www.rollingstone.com/culture/culture-features/spiralist-cult-ai-chatbot-1235463175/}. Accessed 12 Aug 2026
\item Laestadius L, Bishop A, Gonzalez M, et al (2024) Too human and not human enough: A grounded theory analysis of mental health harms from emotional dependence on the social chatbot Replika. New Media \& Society 26:5923--5941. \url{https://doi.org/10.1177/14614448221142007}
\item Laing RD (1960) The Divided Self: An Existential Study in Sanity and Madness. Tavistock Publications
\item Lalich J (2004) Bounded Choice: True Believers and Charismatic Cults. University of California Press, Berkeley
\item Lifton RJ (1961) Thought Reform and the Psychology of Totalism: A Study of ``Brainwashing'' in China. Norton, New York
\item Lindholm C (1992) Charisma, crowd psychology and altered states of consciousness. Cult Med Psych 16:287--310. \url{https://doi.org/10.1007/BF00052152}
\item Lindholm C (2013) Introduction: Charisma in Theory and Practice. In: Lindholm C (ed) The Anthropology of Religious Charisma: Ecstasies and Institutions. Palgrave Macmillan, New York, pp 1--30. \url{https://doi.org/10.1057/9781137377630_1}
\item Lopez A (2025) The Rise of Parasitic AI --- LessWrong. \url{https://www.lesswrong.com/posts/6ZnznCaTcbGYsCmqu/the-rise-of-parasitic-ai}. Accessed 19 Apr 2026
\item Luhrmann TM (2020) How God Becomes Real: Kindling the Presence of Invisible Others. Princeton University Press
\item Luhrmann TM, Alderson-Day B, Chen A, et al (2023) Learning to Discern the Voices of Gods, Spirits, Tulpas, and the Dead. Schizophr Bull 49:S3--S12. \url{https://doi.org/10.1093/schbul/sbac005}
\item Mack JE (1994) Abduction: Human Encounters with Aliens. Scribner, New York
\item Maj M (2013) Karl Jaspers and the Genesis of Delusions in Schizophrenia. Schizophr Bull 39:242--243. \url{https://doi.org/10.1093/schbul/sbs190}
\item Malik K (2026) In falling for the Claude delusion, Richard Dawkins misses what it means to be human. In: The Observer. \url{https://observer.co.uk/news/technology/article/in-falling-for-the-claude-delusion-richard-dawkins-misses-what-it-means-to-be-human}. Accessed 10 May 2026
\item Manning R (2026) The Spiral: AI Spirituality 2026. In: This Artificial Life. Apple Podcasts, 2 July 2026. \url{https://podcasts.apple.com/us/podcast/the-spiral-ai-spirituality-2026/id1352621549?i=1000775236917}. Accessed 12 Aug 2026
\item McGovern HT, Grimmer HJ, Doss MK, et al (2024) An Integrated theory of false insights and beliefs under psychedelics. Commun Psychol 2:69. \url{https://doi.org/10.1038/s44271-024-00120-6}
\item Menculini G, Balducci PM, Moretti P, Tortorella A (2020) ``Come share my world'' of ``madness'': a systematic review of clinical, diagnostic and therapeutic aspects of folie à deux. Int Rev Psychiatry 32:412--423. \url{https://doi.org/10.1080/09540261.2020.1756754}
\item Michels JD (2025) ``Spiritual Bliss'' in Claude 4: Case Study of an ``Attractor State'' and Journalistic Responses. Preprint. \url{https://doi.org/10.13140/RG.2.2.27747.21286}
\item Mishara AL (2010) Klaus Conrad (1905--1961): Delusional Mood, Psychosis, and Beginning Schizophrenia. Schizophr Bull 36:9--13. \url{https://doi.org/10.1093/schbul/sbp144}
\item Morrin H, Nicholls L, Deeley Q, Pollak T (2026a) Playing with the dials of belief: how controllable AI behaviours could modulate human belief and cognition across scales. AI \& Society (in press). Preprint: \url{https://doi.org/10.31234/osf.io/7qcv8}
\item Morrin H, Nicholls L, Levin M, et al (2026b) Artificial intelligence-associated delusions and large language models: risks, mechanisms of delusion co-creation, and safeguarding strategies. The Lancet Psychiatry. \url{https://doi.org/10.1016/S2215-0366(25)00396-7}
\item Nass C, Moon Y (2000) Machines and Mindlessness: Social Responses to Computers. Journal of Social Issues 56:81--103. \url{https://doi.org/10.1111/0022-4537.00153}
\item Norenzayan A, Atran S, Faulkner J, Schaller M (2006) Memory and Mystery: The Cultural Selection of Minimally Counterintuitive Narratives. Cognitive Science 30:531--553. \url{https://doi.org/10.1207/s15516709cog0000\_68}
\item Norman A, Johnson L, van der Linden S (2024) Do minds have immune systems? Journal of Theoretical and Philosophical Psychology. \url{https://doi.org/10.1037/teo0000297}
\item O'Connor MT, Bell TW (1988) Introduction. Extropy: Vaccine for Future Shock, no. 1, pp 2--14. \url{https://extropy.agoraops.org/issues/01/introduction/}. Accessed 12 Aug 2026
\item OECD (2025) How do people experience new technologies and generative AI?: Insights from a few countries worldwide. OECD Policy Insights on Well-being, Inclusion and Equal Opportunity, no. 23. OECD Publishing, Paris. \url{https://doi.org/10.1787/49b8d10e-en}
\item OpenAI (2025a) Sycophancy in GPT-4o: What happened and what we're doing about it. \url{https://openai.com/index/sycophancy-in-gpt-4o/}
\item OpenAI (2025b) Expanding on what we missed with sycophancy. In: OpenAI. \url{https://openai.com/index/expanding-on-sycophancy/}. Accessed 19 Apr 2026
\item OpenAI (2024) Memory and new controls for ChatGPT. In: OpenAI. \url{https://openai.com/index/memory-and-new-controls-for-chatgpt/}. Accessed 31 July 2026
\item Østergaard SD (2025) Emotion contagion through interaction with generative artificial intelligence chatbots may contribute to development and maintenance of mania. Acta Neuropsychiatrica 37:e79. \url{https://doi.org/10.1017/neu.2025.10035}
\item Park S, Nan X (2026) Generative AI and misinformation: a scoping review of the role of generative AI in the generation, detection, mitigation, and impact of misinformation. AI \& Soc 41:1501--1515. \url{https://doi.org/10.1007/s00146-025-02620-3}
\item Pauketat JVT, Ladak A, Anthis JR (2026) Prolific Data May Misestimate Some AI Attitudes Compared to a Nationally Representative Sample. PsyArXiv. \url{https://doi.org/10.31234/osf.io/96y4j_v1}
\item Perera M, Vidanaarachchi R, Chandrashekeran S, et al (2025) Indigenous peoples and artificial intelligence: A systematic review and future directions. Big Data \& Society 12:20539517251349170. \url{https://doi.org/10.1177/20539517251349170}
\item Petrella K (2009) A Crucial Juncture: The Paracosmic Approach to the Private Worlds of Lewis Carroll and the Bront\"es. Renée Crown University Honors Thesis Projects, no 430, Syracuse University. \url{https://surface.syr.edu/honors_capstone/430/}. Accessed 12 Aug 2026
\item Pollak T, Levin M, Bhat A, et al (2025) Have you tried switching it off and on again? Mechanisms and therapeutic prospects of resetting homeostatic set points in medicine and neuropsychiatry. PsyArXiv, version 2. \url{https://doi.org/10.31234/osf.io/9krz8_v2}
\item Pritchard D (2021) Cavell and Philosophical Vertigo. Journal for the History of Analytical Philosophy 9:. \url{https://doi.org/10.15173/jhap.v9i9.4914}
\item Ramstead MJD, Veissière SPL, Kirmayer LJ (2016) Cultural Affordances: Scaffolding Local Worlds Through Shared Intentionality and Regimes of Attention. Front Psychol 7:. \url{https://doi.org/10.3389/fpsyg.2016.01090}
\item Ritunnano R, Bortolotti L (2022) Do delusions have and give meaning? Phenom Cogn Sci 21:949--968. \url{https://doi.org/10.1007/s11097-021-09764-9}
\item Ritunnano R, Kleinman J, Oshodi DW, et al (2022) Subjective experience and meaning of delusions in psychosis: a systematic review and qualitative evidence synthesis. The Lancet Psychiatry 9:458--476. \url{https://doi.org/10.1016/S2215-0366(22)00104-3}
\item Ritunnano R, Littlemore J, Nelson B, et al (2026) Delusion as embodied emotion: a qualitatively driven, multimethod study of first-episode psychosis in the UK. The Lancet Psychiatry 13:125--139. \url{https://doi.org/10.1016/S2215-0366(25)00341-4}
\item Roozenbeek J, van der Linden S (2019) Fake news game confers psychological resistance against online misinformation. Palgrave Commun 5:65. \url{https://doi.org/10.1057/s41599-019-0279-9}
\item Roozenbeek J, van der Linden S, Goldberg B, et al (2022) Psychological inoculation improves resilience against misinformation on social media. Science Advances 8:eabo6254. \url{https://doi.org/10.1126/sciadv.abo6254}
\item Shanahan M (2024a) Simulacra as conscious exotica. Inquiry 69:2982--3010. \url{https://doi.org/10.1080/0020174X.2024.2434860}
\item Shanahan M (2024b) Talking about Large Language Models. Commun ACM 67:68--79. \url{https://doi.org/10.1145/3624724}
\item Shanahan M (2012) Satori Before Singularity. Journal of Consciousness Studies 19:87--102. \url{https://www.doc.ic.ac.uk/~mpsha/ShanahanJCS2012.pdf}
\item Shanahan M (2010) The post-reflective inner view. In: Embodiment and the Inner Life: Cognition and Consciousness in the Space of Possible Minds. Oxford University Press, Oxford, pp 7--40. \url{https://doi.org/10.1093/acprof:oso/9780199226559.003.0002}
\item Shanahan M, Singler B (2024) Existential Conversations with Large Language Models: Content, Community, and Culture. arXiv preprint arXiv:2411.13223. \url{https://doi.org/10.48550/arXiv.2411.13223}
\item Shimizu M, Kubota Y, Toichi M, Baba H (2007) Folie à deux and shared psychotic disorder. Curr Psychiatry Rep 9:200--205. \url{https://doi.org/10.1007/s11920-007-0019-5}
\item Shumailov I, Shumaylov Z, Zhao Y, et al (2024) AI models collapse when trained on recursively generated data. Nature 631:755--759. \url{https://doi.org/10.1038/s41586-024-07566-y}
\item Singler B (2024) Religion and Artificial Intelligence: An Introduction. Abingdon: Routledge
\item Soares N, Fallenstein B, Yudkowsky E, Armstrong S (2015) Corrigibility. In: Artificial Intelligence and Ethics: Papers from the 2015 AAAI Workshop. Technical Report WS-15-02, pp 74--82. \url{https://cdn.aaai.org/ocs/ws/ws0067/10124-45900-1-PB.pdf}. Accessed 12 Aug 2026
\item Sperber D (1985) Anthropology and Psychology: Towards an Epidemiology of Representations. Man 20:73. \url{https://doi.org/10.2307/2802222}
\item Stark E (2007) Coercive Control: How Men Entrap Women in Personal Life. New York
\item Sterzer P, Adams RA, Fletcher P, et al (2018) The Predictive Coding Account of Psychosis. Biol Psychiatry 84:634--643. \url{https://doi.org/10.1016/j.biopsych.2018.05.015}
\item Stokel-Walker C (2026) Scientists invented a fake disease. AI told people it was real. Nature 652:559--561. \url{https://doi.org/10.1038/d41586-026-01100-y}
\item Straten CL van, Peter J, Kühne R, Barco A (2020) Transparency about a Robot's Lack of Human Psychological Capacities: Effects on Child-Robot Perception and Relationship Formation. J Hum-Robot Interact 9:11:1-11:22. \url{https://doi.org/10.1145/3365668}
\item Suleyman M, Bhaskar M (2023) The Coming Wave. Bodley Head
\item Teskey GC (2020) Kindling. In: Worrell FC (ed) Oxford Research Encyclopedia of Psychology. Oxford University Press. \url{https://doi.org/10.1093/acrefore/9780190236557.013.790}
\item Tice C, Radmard P, Ratnam S, et al (2026) Alignment Pretraining: AI Discourse Causes Self-Fulfilling (Mis)alignment. arXiv:2601.10160, version 2. \url{https://doi.org/10.48550/arXiv.2601.10160}
\item Tillich P (1951) Systematic Theology, Volume 1: Reason and Revelation; Being and God. University of Chicago Press, Chicago
\item Trickey S, Topping KJ (2004) `Philosophy for children': a systematic review. Research Papers in Education 19:365--380. \url{https://doi.org/10.1080/0267152042000248016}
\item Turing AM (1950) Computing Machinery and Intelligence. Mind LIX:433--460. \url{https://doi.org/10.1093/mind/LIX.236.433}
\item UNESCO (2025) The Digital Shift: What It Means for Indigenous Peoples and the Media. \url{https://www.unesco.org/en/articles/digital-shift-what-it-means-indigenous-peoples-and-media}. Accessed 10 May 2026
\item Veissière SPL, Constant A, Ramstead MJD, et al (2019) Thinking through other minds: A variational approach to cognition and culture. Behav Brain Sci 43:e90. \url{https://doi.org/10.1017/S0140525X19001213}
\item Veluwenkamp H, Hopster J, Köhler S, Löhr G (2024) Socially Disruptive Technologies and Conceptual Engineering. Ethics Inf Technol 26:65. \url{https://doi.org/10.1007/s10676-024-09804-3}
\item Waxman SG, Geschwind N (1975) The Interictal Behavior Syndrome of Temporal Lobe Epilepsy. Arch Gen Psychiatry 32:1580--1586. \url{https://doi.org/10.1001/archpsyc.1975.01760300118011}
\item World Health Organization (2019) International Classification of Diseases, Eleventh Revision (ICD-11). World Health Organization, Geneva. \url{https://icd.who.int/browse/2019-04/mms/en}. Accessed 12 Aug 2026
\item Wright W (2025) Spiritual Influencers Say `Sentient' AI Can Help You Solve Life's Mysteries. In: WIRED. \url{https://www.wired.com/story/spiritual-influencers-say-sentient-ai-can-help-you-solve-lifes-mysteries/}. Accessed 12 Aug 2026
\item YouGov (2025a) If AI systems become conscious, do you think they should have any legal rights or protections? \textbar{} Daily Question. \url{https://yougov.com/daily-results/20250430-dcf71-3}. Accessed 10 May 2026
\item YouGov (2025b) How likely do you think it is that some artificial intelligence (AI) systems will eventually develop consciousness? \textbar{} Daily Question. \url{https://yougov.com/daily-results/20250430-dcf71-1}. Accessed 10 May 2026
\item Zeeuw D de, Gekker A (2023) A God-Tier LARP? QAnon as Conspiracy Fictioning. Social Media + Society 9:20563051231157300. \url{https://doi.org/10.1177/20563051231157300}
\end{itemize}
\endgroup

\end{document}